\documentclass[useAMS,usenatbib]{mn2e}
\usepackage{psfig,subfigure}

\title[Clumping in H\,II Regions]
{The Effects of Clumping on Derived Abundances in H\,II Regions}

\author[Mathis \& Wood]
{ John S. Mathis$^1$ and Kenneth Wood$^2$, \\
$^1$Astronomy Department, University of Wisconsin, \\
475 N. Charter Street, Madison, WI 53706; mathis@astro.wisc.edu\\
<mailto:mathis@astro.wisc.edu%5C%5C>
<mailto:mathis@astro.wisc.edu%5C%5C>
$^2$School of Physics \& Astronomy, University of St. Andrews,\\
North Haugh, St Andrews, KY16 9SS, Scotland;
kw25@st-andrews.ac.uk <mailto:kw25@st-andrews.ac.uk>
<mailto:kw25@st-andrews.ac.uk>}

\date{Released 2004 Xxxxx XX}
\pagerange{\pageref{firstpage}--\pageref{lastpage}} \pubyear{2004}

\def\LaTeX{L\kern-.36em\raise.3ex\hbox{a}\kern-.15em
T\kern-.1667em\lower.7ex\hbox{E}\kern-.125emX}

\newcommand{\hii}{H\,II}
\newcommand{\np}{N$^+$}
\newcommand{\nepp}{Ne$^{+2}$}
\newcommand{\op}{O$^+$}
\newcommand{\opp}{O$^{+2}$}
\newcommand{\spp}{S$^{+2}$}
\newcommand{\toiii}{$T$([O\,III])}
\newcommand{\tnii}{$T$([N\,II])}
\begin{document}

\label{firstpage}

\maketitle

\begin{abstract}

We have compared Monte Carlo photoionization models of H\,II regions
with a uniform density distribution with models with the same central
stars and chemical compositions but with 3-D hierarchical
density distribution consisting of clumps within clumps, on a
four-tier scheme. The purpose is to compare the abundances of He, N,
O, Ne, and S obtained by standard analyses (emission line strengths
and measured mean temperatures from [O\,III] and [N\,II]) to the
abundances in our models. We consider stellar temperatures in
the range 37.5 -- 45kK and ionizing photon luminosities from 10$^{48}$
-- 10$^{51}$ s$^{-1}$.

Clumped models have different ionic abundances than uniform. For hot
stars, (He$^0$/H$^+$) is 2 -- 3\%, much larger than with uniform
models. This amount of He$^0$ is independent of metallicity and so
impacts the determination of the primordial abundance of He. The total
abundances of O, Ne, and S obtained by the usual methods of analysis,
using \toiii{} for high stages of ionization and \tnii{} for low, are
about as accurate for clumped models as for uniform and within
$\sim$20\% of the true values. If \toiii{} is used for analyzing all
ions, the derived (O/H) is $\sim$ 40 -- 60\% too large for cool stars
but is good for hot stars. Uniform models have similar errors, so the
clumping does not change the accuracy of abundance analysis.

The physical causes of the ionic abundance errors are present in real
nebulae. In clumped models, helium ionizing radiation from zones
of high ionization (low He$^0$ and low UV opacity) can penetrate
nearby regions near the edge of the ionized zone. This effect allows
He$^0$ to absorb more stellar photons than in uniform or radially
symmetrical geometries. In turn, these absorptions compete with
\op, etc., for those energetic stellar photons.

\end{abstract}


\section{Introduction}

The interstellar medium (ISM) has filamentary structure over scales
ranging from hundreds of parsecs through AUs. This paper will analyze
some of the properties of H~II regions formed by ionization of the
clumpy interstellar gas and compare them to those produced by the same
code for models with a spatially uniform density. The Monte Carlo
photoionization code \citep{wme04}
treats gas with densities that have a power-law or scale-independent
character over about an order of magnitude in spatial sizes (see
\S2). We refer to these hierarchical models as ``clumped".

Our goal is to isolate the effects of clumping by comparing clumped
models with smooth density distributions, while keeping other
important parameters (e.g., the stellar atmospheres and luminosities
that strongly influence the nebular models, and the nebular composition)
constant. We choose the other parameters rather arbitrarily and do not
try to compare our models with real objects. Several types of
questions motivate us:

($a.$) For a fixed exciting star and mean nebular density, how are the
ionization and temperature structures of the resulting ionized nebulae
affected? What are the physical causes of the differences? How do the
differences vary as the nebular parameters (e.g., exciting star
properties, geometry of the clumps, and the mean gas density) are
changed?

($b.$) How accurately are model ionic abundances determined from the
emission lines, using techniques similar to those commonly used to
analyze observations of real nebulae? What errors occur in the ionic
abundances derived from ratios of fluxes of emission lines, using mean
nebular electron temperatures, $T_e$, estimated from diagnostic line
ratios?

We assume idealized conditions, neglecting observational errors and
reddening of emission lines by interstellar dust either within or
exterior to the \hii{} region. Thus, our abundance errors are lower
limits to those expected for real nebulae. We consider the effects
of dust in the transfer of Lyman continuum radiation within the
ionized gas.

Our assumptions and the limits of parameter space that we explored are
given in \S2. In \S3\ we consider the physical differences among the
clumped models and their differences from uniform models. In \S4, we
discuss the relevance of our models to real nebulae and the
uncertainties of abundances within them.

We do not address the important issue of differences in abundances of
many ions derived from recombination lines as opposed to collisionally
excited lines. A common explanation is that there are ``anomalous''
temperature variations within the ionized plasma \citep[see, e.g.,
][and references therein]{gr04}, although the phenomenon seems more
complicated \citep[e.g., ][]{tea03}. We have used only standard
physics, and our abundances from recombination lines would be almost
the same as from collisionally excited lines.

\section{Assumptions used in the models}\label{assumpt}
There are many free parameters that affect the spectrum and abundances
derived from model H~II regions, and we consider only a subset of
possible parameter space. We make the following assumptions:

1. The models are in a steady state, thereby neglecting the dynamics
occurring near the outer H$^+ \rightarrow{\rm H}^0$ ionization front,
where probably a D-type front (see, e.g., \citealt{deo89}) is
propagating into the surrounding neutral material. This is the
standard assumption in H~II region models. However, the conditions
near the ionization front are important for N$^0$ and O$^0$, which are
sensitive to the temperature and ionization structure at the
outer edge of the ionized region.

2. We chose the composition used for workshops for testing model H~II
regions (\citealt{fer95}, \citealt{peq01}) : H, He, C, N, O, Ne, and S
= 1, 0.1, 2.2\,10$^{-4}, 4.0\,10^{-5}, 3.3\,10^{-4}, 9.0\,10^{-6}$,
and $5.0\,10^{-5}$, respectively, by number. These abundances are rather
arbitrary, but are adequate for investigating the important physical
effects introduced by clumping. We will refer to elemental abundances
as chemical symbols without superscripts, so that (O/H) is the
abundance ratio of O, including all stages of ionization, relative to H.

3. Dust in \hii{} regions and within the diffuse ionized gas (DIG)
needs careful consideration. We will show below that carbonaceous
grain material has probably largely been converted to the gas
phase. We will find that the effects of carbon-depleted dust are not
very significant, but that with other assumptions dust could become
dominant in determining the structure of the ionized region.

Dust affects the emitted nebular spectrum by
reddening the nebular radiation and also by absorbing ionizing
radiation. Since corrections for reddening
are routinely made, we assume that line strengths are determined
perfectly. This assumption is very optimistic when we consider the
spatially unresolved emission line fluxes rather than the
pixel-by-pixel line intensities (see \S3.2).

If there were no destruction of dust within the ionized zone, the
absorption of ionizing radiation by dust would be hugely important,
while dust scattering, strongly directed forward, has almost no
effect. The absorption cross section of graphitic carbon, closely
related to absorption by polycylic aromatic hydrocarbons (PAHs), peaks
at about 17~eV with a cross section of $3\times10^{-21}$ cm$^2$ (H
atom)$^{-1}$ according to the model of \citet{wd01}, with the cross
sections conveniently tabulated in \citet{d04}. This exceeds the
opacity of H if (H$^0$/H)$\le0.001$, which occurs over most of the
volume of the nebula. Dust with this carbonaceous cross sections
absorbs $\sim$90\% of the ionizing radiation over our entire range of
models, for all geometries. By contrast, the silicate component of
dust is relatively gray in the ionizing UV, with an opacity of roughly
$1.3\times10^{-22}$ cm$^2$ H$^{-1}$, only 5\% of the peak of
carbonaceous dust.

The abundance of dust in \hii{} regions can be estimated by means of
the gas-phase abundances of refractory elements (e.g., C, Al, Mg, Ca,
or Fe; see \citealt{ss96}) that are mostly locked up in dust in the diffuse
ISM. Abundances found in bright \hii{} regions are significantly
different, especially for C. In the diffuse {\em neutral} ISM, the
C\,{\small II} abundances derived from absorption towards 12 B stars
is C/H = 160$\pm$17 ppM and an upper limit of 108 ppM for another
\citep{s04}. In various \hii{} regions, gas phase C/H varies from
330 to 590 ppM (\citealt{eori04}, \citealt{em8}, \citealt{em17},
\citealt{gr04}). Solar (C/H) = 245$\pm$30 ppM (\citealt{a04},
\citealt{hol}). Interstellar abundances have probably been significantly
enhanced since the formation of the Sun, perhaps dex(0.2) to $\sim$400
ppM (e.g., \citealt{aetal04}, \citealt{pt95}). The Orion Bar
\citep{s90}, an ionization front seen edge-on, shows 3.3 $\mu$m 
aromatic hydrocarbon emission arising from a rather thin zone
between the P$\alpha$ and Br$\alpha$ recombination lines of H and
emission lines of H$_2$. Thus, several observations require that bright
\hii{} regions have most or almost all of their carbon in the gas
phase. Usually \hii{} regions have strong near-infrared emission
bands, but it may be that the carriers are actually in the neutral gas
adjacent to the ionized.

In some \hii{} regions, silicaceous dust is present within the ionized
gas. In M\,17, only about 15\% of the Fe is in the gas phase
\citep{em17}. Within the DIG, \citet{hs99} found that Al\,{\small
III}/S\,{\small III} is much less than solar. Both ions are found only
in the gas in which H is ionized, so Al is depleted onto grains in the
DIG. Further direct evidence for dust within the DIG is given by
the correlation of the diffuse H$\alpha$ with 100
$\mu$m emission and the lack of correlation with H\,I \citep{lag00}.

In summary, it seems that the carbonaceous component of dust has
largely been vaporized in \hii{} regions while the silicaceous grains
may largely remain intact. Perhaps processes that convert carbon in
grains to the gas phase act very rapidly in ionized gas, so that the C
in dust (including PAHs and the ``very small grains,'' carbonaceous in
character) does not absorb ionizing radiation when H is mostly
H$^+$. Pure silicate dust has cross sections so low and so weakly
dependent on wavelength that its presence is not crucial to the
results of this paper. We ran models that confirmed this statements,
but for most models in this paper we neglected dust and also heating
by PAHs.

4. Our algorithm for clumping was not derived from any model of
turbulence, but rather from hierarchically clumped models of the
density distributions in the manner described in \citet{mww02},
following \citet{elm97}. Points were placed within a cube as described
below. The cube was partitioned into 65$^3$ cells, and the final gas
density within each of the cells was proportional to the number of
points falling within the cell. The exciting star was placed at the
center of the cube. Initially 32 points were cast randomly throughout
the cube. Surrounding each of these another 32 points were cast
randomly within a distance determined by the ``fractal dimension'',
$D$, of the model (see \citealt{mww02} for details). Around each of
these points another 32 were nested, and finally 32 more around each
of the preceding round of casting. Thus, the structure, if it were
continued indefinitely, would be considered ``fractal'', but the power
spectrum is a power law only over about a decade in sizes of projected
densities. The mean density was taken to be $n$(H) = 100 cm$^{-3}$,
with a fraction $f_{\rm smooth}$ in a uniform distribution and the
rest in the clumped cells. The $f_{\rm smooth}$ represents an average
over a still finer distribution of subclumps, not spatially resolved in
our simulation, that extend down to very small scales.

For this paper we chose $f_{\rm smooth}$ = 0.15, so the minimum
density of a cell is 15 cm$^{-3}$. The maximum density within these
models was 2700 cm$^{-3}$, with a spread of $\pm$15\% arising from
different initial seeds. For comparison to the clumped geometries,
uniform models were computed with the same stellar luminosity and
$n$(H) = 100 cm$^{-3}$. The uniform models fitted more compactly into
the grid cube and consequently could employ a mesh 2.7 times finer
than the clumped models. The abundance patterns found with the coarse
and finer resolution uniform models are in close agreement except for
O$^0$ (the coarse being 1.5 -- 2.2 times larger).

In order to test the effects of allowing radiation to escape from our
clumped models through vacuum instead of unresolved cells, we ran
models with $f_{\rm smooth}$ = 0. About 10\% of the ionizing photons
escaped the simulation, as opposed to $\sim$1.5\% for the $f_{\rm
smooth}$ = 0.15 models.

The distribution of the density in the clumped models is determined by
three parameters: ($a$) $f_{\rm smooth}$, ($b$) the initial seed for the
random number generator, and ($c$) the ``fractal dimension", $D$, that
determines the power spectrum of column densities as projected from an
assumed cube onto the plane of the sky. We considered $D$ = 2.6 and 2.9,
and averaged all clumped models among the same five arbitrarily chosen
initial seeds. Thus, we have an indication of the dispersion of
various physical quantities among clumpy models with the same fractal
dimension.

5. The H-ionizing photon luminosity of the exciting star in units of
10$^{49}$ photons s$^{-1}$, $L_{49}$, was increased from 0.1 to 100,
in steps of factors of ten. For uniform models this is equivalent to
changing the ``ionization parameter", $U$ $\equiv 10^{49}\,L_{49}/
(4\pi R_S^{2} \,n_ec)$, by factors of 10$^{1/3}$ ($R_S$ is the
Str\"omgren radius and $n_e$ the electron density.) A large $U$
implies large abundances of high stages of ionization and a relatively
sharp increase of (H$^0$/H) near the edge of the Str\"omgren
sphere. In clumped models the ionized region is not spherical and
there is no constant value of $n_e$, but $L_{49}$ is an equivalent
parameter to explore the effects represented by $U$ in uniform
models. For instance, our results are applicable to superstar
clusters: $L_{49}\sim10^{3} -10^{4} $, $n_e\sim10^{3} -10^{4} $
cm$^{-3}$ in starburst galaxies.

6. The dielectronic recombination coefficients for sulfur are poorly
known. Unfortunately, they can be important for the ionization of
S. We followed \citet{a91}, using averages of various ions of C, N,
and O. We tested the importance of these assumptions by calculating
some models with all dielectronic recombinations for S being zero. Of
course this is an extreme assumption, since a lack of knowledge of the
values of coefficients does not imply that they are zero. We discuss
the results in
\S\ref{sul}.

The cube for clumped models was large enough so that $\la$1\% of the
ionizing radiation escaped through the edges. The outer boundaries of
the ionized regions were very irregular in shape, and for all stellar
temperatures there are a few regions of neutral H and He embedded
within the ionized gas, with the gas that they shadow from the star
ionized by recombination radiation. 

7. Our cell sizes are large in comparison to the mean free paths of
ionizing photons when H is almost neutral within the cell. Our code
derives a mean ionization for the cell based on the radiation entering
it without considering any substructure of the ionization within the
cell. Our analysis attributes that ionization to the entire mass of
the cell, while only a fraction of the cell's mass would actually be
involved. Near the edge, a few photons can produce ionizations that
the code assumes are spread uniformly over the entire cell, with a
correspondingly low electron density. Therefore, recombinations are
underestimated, and too much mass of ionized material is derived. The
principal ions for which this effect occurs are H$^+$, N$^0$, and
O$^0$. In our analysis we considered only cells in which the code
derived (H$^+$/H) less than H$^0_{\rm min}$, a parameter taken to be
0.25 for the results in this paper. This choice provides roughly an
optical depth of unity across our cells, for a typical photon that can
reach the edge. We have investigated how these results depend the
H$^0_{\rm min}$. Uniform models are hardly affected because the area
of their almost-neutral H zone is much smaller than clumped
models. These show a slow decline in (He$^+$/H$^+$) as H$^0_{\rm min}$
is increased from 0.1 to 0.4, and a more rapid drop from 0.4 to
0.9. The increase of $n$(H$^0$)/$n$(H) between adjacent cells is
typically an order of magnitude when the larger (H$^0$/H) is 0.4 or
more.

We kept the number of cells within the model cube constant and varied
the physical length of the cube proportional to $L_{49}^{1/3}$. This
makes clumped models with the same initial seed have the same
ionization structure in terms of the size of the model cube.

\begin{figure}
\psfig{figure=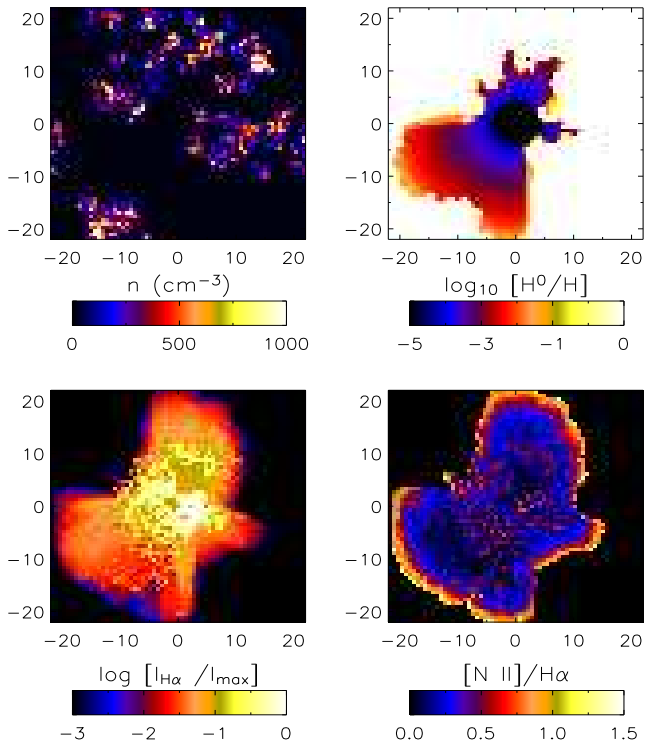,width=4truein,height=4.25truein,angle=0}
\caption{Some properties of the models for $L_{49}$ = 10, $T^*$ = 40 kK.
The axes give distances in parsecs.
{\em Upper left;} A slice through the midplane (containing the star),
showing the densities in the models; {\em upper right:} the fraction
of neutral H; {\em lower left:} a projection of the H$\alpha$ intensity
upon the plane of the sky, normalized to the maximum intensity; {\em
lower right;} the projected [N\,{\small II}]6583/H$\alpha$ ratios. }
\label{panels}
\end{figure}

Fig. \ref{panels} shows ($a$; upper left) the clumped gas densities in
a typical slice through the plane containing the star; ($b$; upper
right) the fraction of H$^0$ in the same slice, ($c$; lower left) the
intensities of H$\alpha$, relative to its maximum intensity, as
projected onto the sky (with no foreground extinction), and ($d$;
lower right) the [N\,{\small II}]/H$\alpha$ ratio projected onto the
sky. The axes are in parsecs for a model with $T^*$ = 40 kK, $L_{49}$ =
10. We see that the density through the midplane is concentrated to
the left side, but the projected images, ($c$) and ($d$), are more
symmetrical because of ionization in the foreground and background
planes. The [N\,{\small II}]/H$\alpha$ is large at the edge because
the ionization favors N$^+$ and the temperature is relatively high
there.

The large [N\,{\small II}]/H$\alpha$ (lower right panel) at the edges
does not imply that the \hii{} region would appear limb-brightened in
the sky. The lower left panel shows that the H$\alpha$ intensity at
the edges is only $\sim$1\% as bright as the maximum, so the
[N\,{\small II}] intensity is low and would blend in with the
unrelated surrounding diffuse emission. It is only the [N\,{\small
II}]/H$\alpha$ {\em ratio} that increases at the edges.
\section{Effects of clumping on ionic abundances}
Our results from a large number of models can be summarized simply:
for a given exciting star, clumping changes the averaged fraction of
ionization of all elements, but differently for relatively cool stars
than for hot. The important quantity is $\gamma \equiv (L_{\rm
He}/L_{\rm H})/$(He/H). If $\gamma>1$ a star is ``hot''; $<1$,
``cool.'' Clumping enhances the higher stages of ionization of an
element (including He) for cool stars and decreases them for hot. The
amounts of enhancement for specific ions are discussed below.

We have $L_{\rm He}=\int L_\nu^*/h\nu\,d\nu$, integrated from the
ionization potential of He$^0$, $\chi$(He$^0$) = 24.6 eV, to
$\chi$(He$^+$) = 54 eV. $L_{\rm H}$ is $\int L_\nu^*/h\nu\,d\nu$, 
integrated from 13.6 eV to 54 eV. We count the radiation above 24.6 eV
as both H- and He-ionizing because each recombination of He produces
about one H-ionizing photon \citep{deo89}, so absorptions by He hardly
affect the number of H-ionizing photons. Interstellar dust does not
affect the distribution of ionization significantly if it contains
only silicates, as seems likely (see \S\ref{assumpt}).

\subsection{Influences of parameters}
There are a large number of parameters in our models. Some set the
basic physics and radiative transfer (elemental abundances, $T^*$,
$L_{49}$, and $f_{\rm smooth}$), but others are arbitrary: the fractal
dimension, $D$, and the initial seed that determines the density
structure in clumped models. The dispersion of the
results among the five initial seeds that we arbitrarily chose for
each clumped model is not large. Typical variations of (H$^+$/H) are
0.1\%; of (He$^0$/He), 0.2\% if (He$^0$/He) $\le$ 0.1 and $<$2\% if
(He$^0$/He)$\sim$ 0.3. The mean temperatures weighted by
$n_e\,n($H$^+$) vary $\sim$0.5\%. Varying $D$ from 2.6 to 2.9 produced
changes about twice the dispersion among the various seeds of each
clumped model. The overall result is that the other parameters were
much more influential than $D$ and the initial seed. For the rest of
the paper, we will present $D$ = 2.6 results as representing clumped
models, with results averaged among the five seeds.

We begin by considering the main physical difference between clumped
and uniform models, then define averaged abundances for comparing
models, and then consider specific elements and ions.

\subsection{Propagation of He Ionizing Radiation \label{physdif}}
The most important consequence of a clumped density distribution is
the fate of He ionizing radiation, depending upon the hardness of the
stellar radiation. It determines the relative amounts of the nebular
opacity supplied by H and He, the two big players for almost
all ionizing photons.

Hot stars produce bountiful He-ionizing photons, so H can compete with
He for He-ionizing radiation. At $h\nu$ = 25 eV, just above
$\chi$(He$^0$), H contributes $\sim$40\% of the opacity near the star
with $T^*$ = 45 kK, and He practically all of the rest. In a uniform
model, the fraction of H opacity gradually falls with distance from
the star as He becomes more neutral. In the interclump regions of
clumped models, the H opacity decreases less rapidly with distance
because the density is lower. The abrupt increase of density when
stellar photons encounter a clump makes the increase of He opacity
over H occur within a small volume. The overall effect is to make H
compete for He-ionizing radiation more effectively in clumped models
than in smooth, thereby decreasing (He$^+$/He) and other high stages
of ionization over uniform models.

He supplies 95\% of the opacity for 25 eV photons when $T^*$ = 37.5
kK.  In this case, in clumped models the He is ionized in the
interclump medium until the outflowing starlight encounters a clump,
at which point both H and He tend to become neutral. In a uniform
model, the He goes neutral after it absorbs essentially all of the
stellar He-ionizing radiation. and there is a substantial region in
which H is still H$^+$ but He is neutral.  The overall effect is that
He and H tend to be both ionized within the same volume in clumped
models, so the (He+/He) within the ionized region is larger than in
uniform models. The same effect applies to other high stages of
ionization. {\em This is the major physical cause of differences
between clumpy and smooth models.}

We compared clumped models with those with uniform densities, but the
root physical causes of the differences lie in the radial symmetry,
not the uniformity, of the uniform models. The general distinctions
between clumped and uniform models applies to any smooth distribution
of gas density.

We considered \citet{k94} stellar atmospheres of (37.5, 40, 42.5, and
45) kK, log $g$ = 4 and used a power-law interpolation between them to
estimate spectra for 38.1, 38.7, and 41 kK as well. For these atmospheres,
$\gamma =1$ occurs at about 39.1 kK. We will see qualitative
differences in the ionization structure for other elements at this
$T^*$. For Kurucz atmospheres, $\gamma$ = (0.569, 1.35, 2.41) for
$T^*$ = (37.5, 40, 45) kK. We present plots of ionization fractions vs.
$T^*$ because $\gamma$, the real independent variable, is a less
familiar quantity.

\subsection{Comparing Models with Observables}
Our aim is to compare the abundances within our clumped models with
the abundances that would be inferred if the emissions from the
models, measured without error or dust, were interpreted in the ways
used on actual nebulae. The derived abundances differ from the actual
ones in the model because of temperature and density variations within
the model, even if the density is never enough to de-excite the
emission lines by collisions. If there is poor spatial resolution,
there is averaging over the face of the nebula as well. We will
compare three types of averaged abundances of a particular ion, $X^i$,
relative to H. The line emission (photons
cm$^{-3}$\,s$^{-1}$\,ster$^{-1}$) is $n(X^I)n_e \,j_\lambda(X^i,T_e)$,
and $j_\lambda$ is presumed to be known from atomic theory.

Since the model provides 3-D ionization structure of the nebula, the
true abundances are known:

\begin{equation}
\left(\frac{X^i}{{\rm H}^+}\right)_{\!\!\rm true}= \frac{\int n({\rm
He}^+)\,dV}
{\int n({\rm H}^+)\,dV} \label{true}\ .
\end{equation}

An observer would know the projected line intensities, $I_\lambda =
\int\, n(X^i)n_e\,j_\lambda(X^i,T_e)\,dz$, where $z$ is the distance
along the line of sight, averaged over the observer's spatial
resolution. An extreme assumption would be that there is no spatial
resolution over the face of the nebula, so that we know only the total
line fluxes, $F_\lambda$, thereby obtaining a ``global'' averaged
abundance:
\begin{eqnarray}
F_\lambda&=&\int I_\lambda \,dx\,dy\ ,\\
&=&\int n(X^i)n_e\,j_\lambda(X^i,T)\,dV\ . \\
\left(\frac{X^i}{\rm H^+}\right)_{\!{\rm global}}&
=&\frac{{\normalsize F_\lambda(X^i)/
j_\lambda(X^i,T_X)}}{{\normalsize F(H)/j_H(T_H)}} \\
&=& \frac{\int
n(X^i)\,w(X^i)\,dV}
{\int n(H^+)\,w(H)\,dV} , \label{global}
\end{eqnarray}
\noindent where $T_X$ is the measured averaged temperature relevant to
the ion $X^i$, depending upon whether it is a high or low stage. The
weights $w(X^i)=n_e\,j_\lambda(X^i,T)$, and similarly for the H
lines. For high ions we use \toiii, obtained from [O\,III]$\lambda
$4363/(5007+4859). For low ions we use \tnii, from
[N\,II]$\lambda$5755/(6583+6548).

We will explore the effects of the $w(X^i)$ on the derived abundances.
Note that the effect is not limited to variations in temperature: it
exists even in artificial isothermal models because of variations in
$n_e$.

Another extreme assumption is that ($X^i$/H) can be determined at each
point ($x,y$), and the result, $A_X(x,y)$, then averaged, weighting
the bright pixels more than the faint. We define the abundances
averaged over the face of the nebula by
\begin{equation}
\left( \frac{X^i}{H^+}\right)_{\!\rm ave} = \frac{\int A_X(x,y)\,
I_H(x,y) \,dx\,dy}{\int I_H(x,y) \,dx\,dy}\ . \label{ave}
\end{equation}

The fluctuations of temperature within our models are modest, as
uniform and smooth models have previously suggested. The variation of
\toiii{} across the face of a nebula \citep[e.g., ][]{opp03} is
described by $t^{2} =\int (T(x,y)-T_0)^{2} dx\,dy/[T_0^{2} \,\int dx\,dy$],
where $dx\,dy$ is the element of area on the sky and $T_0$ is the
averaged temperature. \citet{opp03} found $t^{2} $ in the range 0.005 --
0.016 in five regions of the face of the Orion Nebula using HST images
of $\lambda$4363 and 5007 with high spatial resolution. For $T^*$ =
40kK and $L_{49}$ = 10, our models collapsed along one axis yield $t^{2}
$ = 0.0012 for the uniform model and $(1.0-1.6)\times10^{-3}$ for the
clumped models with various initial seeds. For $T^*$ = 45 kK, $t^{2} \sim
0.002$. Thus, the Orion Nebula shows temperature fluctuations in
excess of models, but not enough to explain the discrepancies between
collisionally excited and recombination lines unless one assumes that
in real nebulae more violent fluctuations are averaged out by the
integration along the line of sight.

We do not discuss fine structure lines ([O\,III] 52, 88 $\mu$m, plus
several others) because their emissivities are not sensitive to
variations in temperature or densities within the rather modest ranges
that we consider. These lines provide excellent diagnostics of
the true abundances of their respective ions for both clumped and
uniform models, but are subject to collisional de-excitations in
real objects.

We now discuss individual ions and elemental abundances.

\subsection{{\bf {\rm He}$^+${\rm /H}$^+$}}

\begin{figure}
\psfig{figure=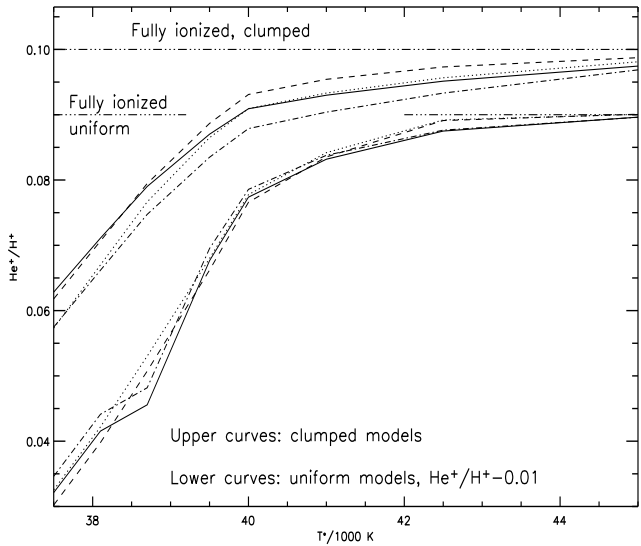,width=3.5truein,height=3.0truein,angle=90}
\caption{The ratio (He$^+$/H$^+$) as determined for clumped (upper
set of curves) and uniform models, plotted against the exciting star
temperature, $T^*$. The uniform models show [(He$^+$/H$^+$)-0.01]. The
horizontal lines show complete ionization of He for clumped and
uniform models. Solid line are true abundances (see eqn. \ref{true})
for stars with $L_{49} = 10^{-49}L_*$ = 0.1; $L_*$ is the ionizing
photon luminosity. Dashes: true abundances for $L_{49}$ =
100. Dot-dashes: global clumped abundances that would be derived by
observers (see eqn. \ref{global}) for $L_{49}$ = 0.1; dotted, global
abundances for $L_{49}$ = 100. The ratio of stellar ionizing photon
luminosities for He and H, $L_{\rm He}/L_{\rm H}$ equals the elemental
abundance ratio (He/H) at $T^*$ = 39.5 kK, producing the bends in the
curves. }
\label{heh}
\end{figure}

Figure \ref{heh} displays (He$^+$/H$^+$) plotted against $T^*$ for
clumped models (upper sets of curves) and (He$^+$/H$^+-0.1$) for
uniform density distributions (lower sets). The horizontal lines at
(He$^+$/H$^+$) = 0.1 and 0.09 show the cases of complete ionization of
He. The solid lines are true abundances (eqn. \ref{true}) for $L_{49}$
=0.1; the dashed lines are for $L_{49}$ = 100. The dot-dashed lines
are the global abundances (that would be deduced by observers using
the fluxes of lines) for $L_{49}$ = 0.1, the dotted for $L_{49}$ =
100. The other values of $L_{49}$ have curves lying between the ones
shown. The bends in the curves near $T^*$ = 39.5 kK occur because
$\gamma\sim1$, so that the ratio of stellar ionizing photon
luminosities for He and H, $L_{\rm He}/L_{\rm H}$ equals the elemental
abundance ratio (He/H).

Fig. \ref{heh} shows that clumped models provide a larger He$^+$/H$^+$
at a given low $T^*$ than uniform models; $\sim$0.06 at $T^*$ = 37.5
kK, as opposed to $\sim0.04$ for uniform. For hot stars, the uniform
models (the lower set of curves) have (He$^+$/H$^+$) = (He/H) to high
accuracy, but this is not so for clumped nebulae. They have
$\sim$2.5\% of He$^0$ even for $T^*$ = 45 kK for the true abundances,
and the derived global (He$^+$/H$^+$) (i.e., weighted by $n_e$) is
98.1\% of (He/H). The difference between the derived (He$^+$/H$^+$)
and (He/H) is 3.8\% for 42.5kK stars. The error increases for lower
stellar temperatures and $L_{49}$.

The relatively large fraction of neutral He for clumpy nebulae
surrounding hot stars has important consequences:

(a) (He/H) is a major prediction of Big Bang Nucleosynthesis, and the
comparison with real nebulae requires very high accuracy in the
correction for He$^0$ (see \citealt{os04} for a review showing the
accuracy required). The reason why clumped models exhibit more He$^0$
than uniform models is independent of metallicity. The observations
and interpretation of He\,I emission lines in metal-poor extragalactic
H~II regions strive for accuracies of $\sim$1\% in He$^+$/H$^+$
\citep[e.g., ][]{ppm00}. There must be corrections for stellar
absorption lines underlying the nebular emissions as well as density
and temperature effects on the He\,I line strengths. Uniform or smooth
models for nebulae ionized by collections of very hot stars predict
that He$^+$/H$^+$ = He/H. Clumped models suggest an increase of He/H
of the order of $\sim$ 3\% to correct for He$^0$. Equation 11 of
\citet{os04} suggests that uncertainties in the determination of the
primordial He$^+$/H$^+$ amount to $\sim$10\%, but the correction for
He$^0$ is systematically upwards.

(b) Helium is dominant in determining the opacity for energetic
photons that produce ions of other elements, exceeding hydrogen by a
factor of a few near its ionization edge. For harder radiation
($h\nu>\chi$(O$^+$) = 35 eV), its main competitor, O$^+$, typically
provides $<$30\% of the opacity, and the other ions contribute
significantly less because of their relatively low elemental
abundances. Thus, the increased He$^0$ decreases the abundances of
high stages of ionization of heavy elements.

\subsection{Heavy Elements}\label{hel}
For heavy elements, the ionic abundances are strongly influenced by
whether He recombinations to the ground term, $h\nu\sim25-26$~eV, can
ionize the stage in question (e.g., $\chi$(S$^+$) = 23.3~eV), so that
the He recombination radiation competes with stellar radiation. If the
He radiation cannot produce ionization, for cool stars the lower stage
(\op, Ne$^+$, N$^+$) is reduced in clumped models as compared to
uniform. This behavior is similar to He$^0$ (figure \ref{heh}) because
lower He$^0$ implies a smaller opacity for hard stellar photons.

There are systematics in the temperatures measured in various
ways. For all geometries, $T$([O\,III]) is cooler than the
$n_e$-averaged temperature, $\langle T\rangle\equiv \int
T\,n_e\,dV/\int n_e\,dV$, by $\sim$300 K because \opp{} is a powerful
coolant. \tnii{} is warmer than $\langle T\rangle$ by $\sim$400 K
because stellar photons are always harder on average at the outer
regions of the ionized gas where [N\,II] is produced, and \op{} is
not as efficient a coolant as \opp
. Clumped models
have warmer $\langle T\rangle$ than uniform; for $T^*$ = 37.5 kK, the
difference is 50 -- 100 K; for $T^*$ = 45 kK, it is $\sim180$
K. \tnii{} for uniform models is about 50 K larger than for
clumped. These temperature differences affect the derived abundances.

\subsubsection{High stages of Ionization: {\rm O}$^{+2}$ and
{\rm Ne}$^{+2}$} This section deals with the two ions with very
similar behavior as regards differences between uniform and clumped
models: \opp{} and \nepp. Their important characteristics are that
they cannot be ionized by He recombination radiation and cannot be
ionized to still higher stages by stellar radiation ($\chi$(\opp) =
54.9 eV; $\chi$(\nepp) = 63.4 eV). This group does not include \spp{}
($\chi$ = 34.8 eV) because it can be ionized to S$^{+3}$ by stellar
radiation. We will assume that there are only observations of N$^+$
optical lines and discuss nitrogen in the next section, although there
are fine-structure transitions of N$^{+2}$ at 57 $\mu$m and of
N$^{+3}$ at 69 $\mu$m.

\begin{figure}
\psfig{figure=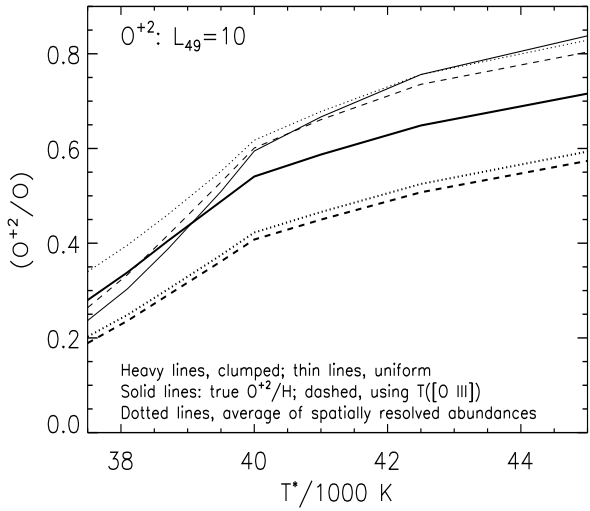,width=3.5truein,height=3.0truein,angle=90}
\caption{The fraction of oxygen in \opp{} vs. $T^*$ for
$L_{49}$, the exciting stellar photon luminosity in units of 10$^{49}$
sec$^{-1}$, =10. The heavy lines are clumped models; the thin lines,
uniform. The solid lines are the true abundances; the dashed are
derived from \toiii; the dotted are the spatially resolved abundances,
averaged over the face of the nebula (see equation \ref{ave}.)
}
\label{oppts}
\end{figure}

Figure \ref{oppts} shows (\opp/O) plotted against $T^*$ for both true
abundances and the global averages, for $L_{49}$ = 10. Heavy lines are
clumped models and thin are uniform. The true abundances are solid
lines and the global abundances using \toiii{} are dashed.

\begin{figure}
\psfig{figure=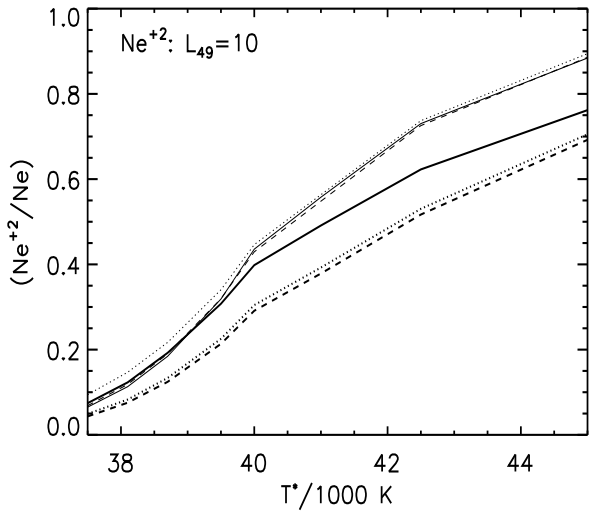,width=3.5truein,height=3.0truein,angle=90}
\caption{Plots of the (\nepp/Ne) ratio, with the same notation
as used for figure \ref{oppts}. Note the strong similarities, for
reasons discussed in the text. }
\label{nep2ts}
\end{figure}

Figure \ref{nep2ts}, showing (\nepp/Ne), is analogous to
Figure \ref{oppts}. The similarities in the ionizations are
clear. \nepp{} starts lower than \opp{} at low $T^*$ but increases
faster because it requires harder radiation: $\chi$(Ne$^+$) = 40.96
eV; $\chi$(\op) = 35.1 eV.

Figures \ref{oppts} and \ref{nep2ts} show that:

1. For clumped models the accuracy of the global (\opp/O) abundances derived
from \toiii{} is only $\sim$70 -- 80\% of the
true. {\em Even perfectly accurate [O\,III]
and H$\beta$ line intensities do not lead to an accurate determination
of \opp{} abundances.} The reason is that there are relatively strong
temperature fluctuations within the \opp{} zone in clumped models.
Excess [O\,III] $\lambda$4363 is produced by the warmer regions,
making the mean \toiii{} within the zone higher than the true mean
temperature. This situation leads to an erroneously low abundance
being determined for \opp. (\nepp/Ne) is somewhat more accurately
determined than (\opp/O).

For uniform models, \toiii{} global abundances are in excellent
agreement with the true. Surprisingly, the accuracy of the \toiii{}
abundances is better for \nepp{} than for \opp. For both uniform and
clumped models, (\nepp/Ne)$\sim$(\opp/O) until both are 

2. The averaged spatially resolved abundances (dotted lines) are
almost the same as those obtained with no spatial resolution.

3. The true (\opp/O) and (\nepp/Ne) are larger for clumped models than
for uniform in cool stars and smaller for hot stars, as was true for
(He$^+$/H$^+$) (see Figure
\ref{heh}). 

The behavior of (\opp/O) and (\nepp/Ne) at other values of $L_{49}$ is
as expected. Decreasing $L_{49}$ decreases high stages of ionization:
for $L_{49}$ = 0.1, the set of curves has the same form as those
already plotted but lie entirely under the envelope of the curves
shown.

\subsubsection{Low Stages of Ionization: {\rm O}$^+$ and {\rm N}$^+$}
These ions and their next higher stages have far larger abundances
than the remaining stages of ionization (O$^0$, N$^0$, and N$^{+3}$)
over the bulk of the nebula. Both \op{} and \np{} are further ionized
only by stellar radiation: $\chi$(\op) = 35.12 eV and $\chi$(N$^+$) =
29.6 eV, well above the energy of almost all He
recombinations. 

\begin{figure}
\psfig{figure=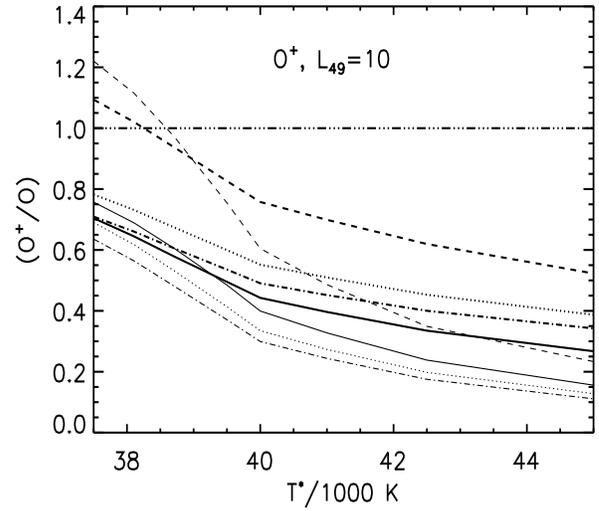,width=3.5truein,height=3.0truein,angle=90}
\caption{Global (\op/O) vs. $T^*$ for $L_{49}$ =
10. The heavy lines are clumped models; the thin, uniform. The dashed
lines are derived from \toiii; the dot-dashed from \tnii. The dotted
lines are averaged abundances.
}
\label{opts}
\end{figure}

\begin{figure}
\psfig{figure=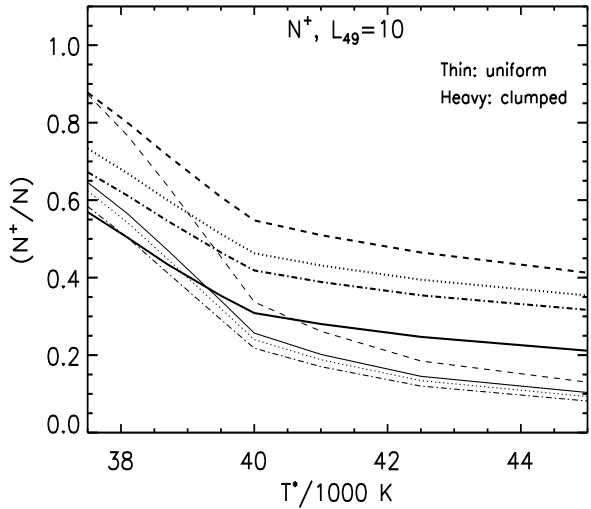,width=3.5truein,height=3.0truein,angle=90}
\caption{Plots of the \np{} against $T^*$ for $L_{49}$ = 10 for
clumped models (heavy lines) and uniform models (thin lines). Solid
lines are true abundances. Dot-dashes: global abundances derived from
\tnii. Dashes: global abundances using \toiii. Dots:
average abundances (see equation \ref{ave}).
}
\label{npts}
\end{figure}

The (\op{}/O) global ratios for $L_{49}$ = 10 are shown in Figure
\ref{opts}. Heavy lines are clumped models, thin are uniform.
Solid lines are true abundances; dashed are global abundances derived
from \toiii{} and dot-dashed from \tnii. The dotted lines are averaged
abundances. Figure \ref{npts} gives (\np/N). Both show that the use
of \toiii{} for a low stage of ionization is problematic, especially
for cool stars. The \tnii{} abundances for N$^+$ are
$\sim$40\% too large but are, surprisingly, more accurate for (\op/O)
than for (N$^+$/N). 

Using \toiii{} for \op{} and \np{} is especially
problematic for clumped nebulae with hot exciting stars; the
abundances will likely be overestimated by a factor of
$\sim$1.5. Using \tnii{} provides results within $\sim$20\% of the
true for (\op/O) and $\sim$30\% for (\np/N).

\subsubsection{Neutrals: {\rm O}$^0$ and {\rm N}$^0$}
In regions where these ions are abundant, their ratio to the
singly-charged species is almost completely determined by charge
exchange with H$^0$. [O\,I] $\lambda$6300 emission is produced almost
entirely in a physically thin zone at the outer edge of the H$^+$
zone. In this region there is appreciable H$^0$ ($\ga$0.5\%) for
charge exchange and a reasonably high $T_e$, needed to produce the
collisional excitation of [O\,I]. Newly ionized gas flows away from
the neutral material, and gas dynamics, neglected in our models, plays
an important role in the local temperature and ionization
structure. Both the width of the zone affected by dynamics and the
mean free path of ionizing photons are less than our
cell size, so we resolve this region poorly. We do not present
predictions for (O$^0$/O), which we find to be $\sim$3\% for $L_{49}$
= 0.1 and 1\% for $L_{49}$ = 100. These values are almost surely
overestimates. Our (N$^0$/N) is similarly unreliable. The cells in
which 0.25$\le$(H$^+$/H)$\le$0.9 have a mass of $\sim$8\% of the
highly ionized plasma in uniform models and $\sim$30\% in
clumped. However, the temperature in the partially ionized gas is
falling rapidly as H$^+$/H decreases, so the emissions from the almost
neutral gas are small. The region $0.9< ({\rm H}^+)/{\rm H}<0.99$ adds
only $\sim$1\% to the mass and is quite cool ($<$6000 K), so it
contributes a negligible amount of emission.

\subsubsection{Sulfur\label{sul}}
Sulfur has four relevant stages of ionization: S$^+$ ($\chi$ =
23.3~eV, so it is ionized by the most energetic He recombination
radiation); \spp{} ($\chi$ =34.8~eV, similar to \opp); S$^{+3}$
($\chi$ =47.3~eV), and S$^{+4}$. There is too little stellar radiation
to make S$^{+4}$ important in any of our models.

\begin{figure}
\psfig{figure=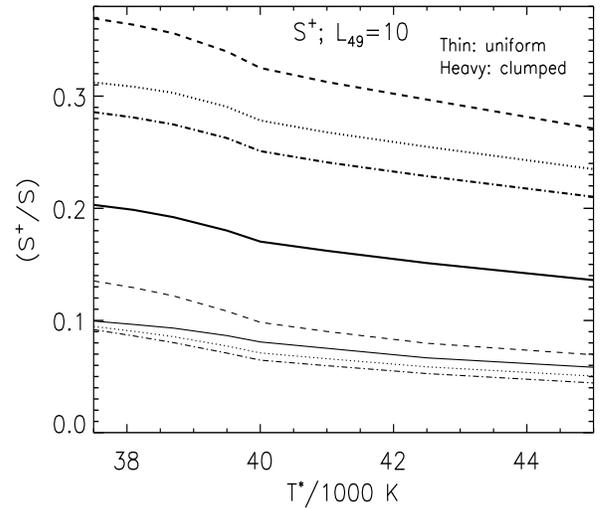,width=3.5truein,height=3.0truein,angle=90}
\caption{Plots of the (S$^+$/S) against $T^*$ for $L_{49}$ = 10 for
clumped models (heavy lines) and uniform models (thin lines). Solid
lines are true abundances; dot-dashes, global abundances derived from
\tnii; dashes, global abundances using \toiii; dots,
average abundances (see equation \ref{ave}). Compared with Figure
\ref{opts}, the same diagram for \op, we see the influence of the
ionization of S$^+$ by He recombination radiation that does not affect
\op.
}
\label{spts}
\end{figure}

\begin{figure}
\psfig{figure=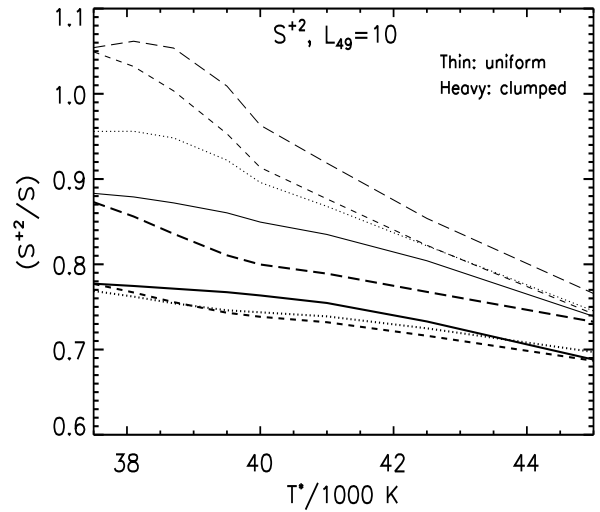,width=3.5truein,height=3.0truein,angle=90}
\caption{Plots of (\spp/S) against $T^*$ for $L_{49}$ = 10 for
clumped models (heavy lines) and uniform models (thin lines). Solid
lines are true abundances; dashes, global abundances using \toiii; dots,
average abundances (see equation \ref{ave}).
}
\label{sppts}
\end{figure}

Figures \ref{spts} and \ref{sppts} show (S$^+$/S) and (\spp{}/S) plotted
against $T^*$, both for $L_{49}$ = 10. The scheme of the lines is the
same as in Figures \ref{oppts} -- \ref{npts}. These ions do not have
the properties of the corresponding stages of oxygen (Figures
\ref{oppts} and \ref{opts}).

Comparison of Figure \ref{spts} with Figure \ref{opts} reveals that
S$^+$ has a much smaller true fractional abundance than \op{} because
He recombinations ionize S$^+$. (S$^+$/S) varies far less with $T^*$
as well. A greater fraction of O is in \op{} than S in S$^+$, so the
clumpy nature of clumped models makes more difference for S$^+$. The
abundance of S$^+$ in clumped models is three or more times larger
than for uniform.

For clumped models, the global abundance of S$^+$ derived from
\toiii{} (the dashed lines in Fig. \ref{spts}) is
about twice the true abundances. The S$^+$ derived from \tnii{} is
about 50\% too large.

Figure \ref{sppts} shows the trends of (\spp/S). The bulk of sulfur is
\spp, but both S$^+$ and \spp{} decrease with increasing $T^*$,
showing the growing importance of S$^{+3}$. 

The physically implausible assumption that all dielectronic
recombination coefficients for S are zero leads to an appreciable
shift of the abundances of S ions to the higher stages: from (S$^+$:
S$^{+2}$: S$^{+3}$: S$^{+4}$) = (0.172, 0.762, 0.066, $5\times
10^{-5}$) for $L_{49}$ = 10, $T^*$ = 40 kK clumped models to (0.115:
0.702: 0.182: 0.003) with no dielectronic recombination for S. Besides
the obvious direct consequences for the S line strengths, the averaged
electron temperature increases from 8751 K to 8812 K because S$^{+2}$
is a major coolant. However, the effects of geometry, $T^*$, and
$L_{49}$ dominate the effects of the uncertainties in the dielectronic
recombination coefficients of S.

\subsubsection{Total abundances: O, Ne, and S}
Assuming that we have measures of both \toiii{} and \tnii, how well
can we estimate total elemental abundances from normally observed
emission lines? Oxygen might provide the best case, since all of its
ionization stages are observable. We include our estimates of O$^0$ in
our discussion. In a real nebula it would be smaller than our
prediction, while O$^+$ would be correspondingly larger.

\begin{figure}
\psfig{figure=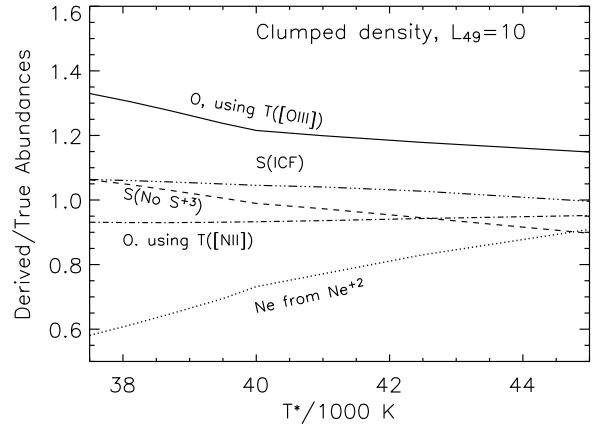,width=3.5truein,height=2.5truein,angle=90}
\caption{The total estimated global abundances of various elements.
The elements and method of estimating its abundance are noted by the
lines. For oxygen, ``using \toiii'' refers to the temperature used for
\op{} and O$^0$, and similarly for ``using \tnii''. \toiii{} is always
used for \opp. The S(ICF) is from an Ionization Correction Factor
to account for S$^{+3}$ (see text).}
\label{ftotts}
\end{figure}

Figure \ref{ftotts} shows the estimated global abundances of O, Ne,
and S for clumped models, relative to the true abundance. The
estimates for oxygen indicate which temperature is used to estimate
\op. \toiii{} is always used for \opp. The estimate for Ne is derived
from scaling the Ne$^{+2}$ abundance by the model predictions of
(Ne/Ne$^{+2})$. The line marked ``S(no S$^{+3}$)'' simply ignores that
ion. The S(ICF) is based on an empirical Ionization Correction Factor
(ICF) to account for S$^{+3}$ (\citealt{kw01}). It relates
S/(S$^++{\rm S}^{+2}$) to (O/\op):
\begin{eqnarray}
{\rm S}&=&({\rm S}^++{\rm S}^{+2})\times ICF\ ;\\
\log (ICF)&=&-0.17+0.18\beta-0.11\beta^{2} +0.072\beta^{3} \ ; \\
\beta&=&\log ({\rm O}/{\rm O}^+)\ .
\end{eqnarray}

\begin{figure}
\psfig{figure=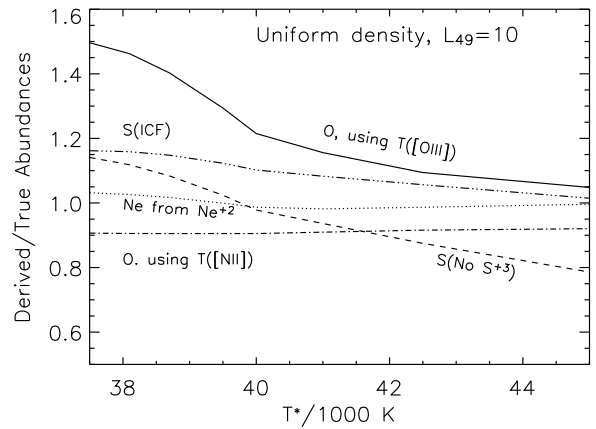,width=3.5truein,height=2.5truein,angle=90}

\caption{The total estimated global abundances of various elements.
The elements and method of estimating its abundance are noted by the
lines. For oxygen, ``using \toiii'' refers to the temperature used for
\op{} and O$^0$, and similarly for ``using \tnii''. \toiii{} is always
used for \opp. The S(ICF) is from an Ionization Correction Factor
to account for S$^{+3}$ }
\label{utotts}
\end{figure}
\noindent We see that the ICF provides an excellent estimate for the
abundance of S.

Figure \ref{utotts} is similar to Figure \ref{ftotts}, except for
uniform densities. For clumped models there are compensating
errors when we use \tnii{} for \op{} and \toiii{} for \opp: the \op{}
is too large; the \opp{} too small. For uniform models, both are only
slightly too small. The net result is that the estimated abundance
(the dot-dashed lines in figures \ref{utotts} and \ref{ftotts}) is
found to be $\sim310$ ppM for all $T^*$ and $L_{49}$, both uniform and
clumped. Recall that in our models the true (O/H) = 330 ppM.

The importance of being able to use a suitable mean temperature [say,
\tnii] for low excitation ions is clear, especially for cool stars. If
we are forced to use \toiii{} for \op{} as well as \opp, both
estimates are almost always too large (especially for \op; see
Fig. \ref{opts}, dashes). They average about 400 ppM, and the worst
(uniform, cool stars, $L_{49}$ = 100) is a factor of 1.7 too
large. The best case is for uniform geometry models excited by hot
stars with $L_{49}$ = 100. Its predicted oxygen is almost correct, but
the corresponding clumped models predict 20\% more than the correct
value. Clumping introduces appreciable uncertainties.

The plots of total abundances against $T^*$ for other values of
$L_{49}$ are quite similar to those shown in Figures \ref{ftotts} and
\ref{utotts} for $L_{49}$ = 10. For $L_{49}$ = 0.1, the O using
\toiii{} for \op{} (the solid line) is relatively flat at $\sim$120\%
for both clumped and uniform models, an improvement over the
situation at $L_{49}$ = 10. For $L_{49}$ = 1, the curves are close to
those in the figures. The total S without S$^{+3}$ is also better for
lower $L_{49}$, since the levels of high stages of ionization increase
with $L_{49}$. For $L_{49}$ = 100, the total S without S$^{+3}$ varies
from 110\% of the true at $T^*$ = 37.5 kK to 83\% at 45 kK for
clumped and 120\% to 66\% for uniform models.

\section{Remarks and Summary}
The clumped ionization fractions differ significantly from uniform, and
there are significant errors in estimating ionic abundances from
emission line fluxes. (He$^+$/H$^+$) is significantly increased for
cool stars by clumping,
and (He$^0$/He) is appreciable (2 -- 3\%) for even very hot stars.
Clumped models show less (\opp/O) and (\nepp/Ne) than uniform for hot
stars and more for cool. For clumped, the abundances of both ions
estimated by the standard methods using emission lines and the
appropriate mean temperature, \toiii, is only $\sim$70\% of the true
abundances. For uniform models, the derived ionic abundances agree well
with true. Real nebulae probably show the same phenomenon, and no
improvement in spatial resolution will resolve it.

The strong effects of clumping on ionic abundances for a given stellar
energy distribution (\S\ref{hel}) makes the determination of the
properties of the exciting star from the nebular ionic ratios even
more problematic than it is with smooth models. Another uncertainty is
the presence of carbonaceous dust, with a strong absorption at
$\sim$17 eV (\S\ref{assumpt}), that makes the spectrum of the
exciting star appear hotter because H-ionizing radiation is absorbed
more readily than He-ionizing. Silicate dust makes little difference
to \hii{} regions because it has a low, almost wavelength-independent
cross section. 

Real H\,II regions have irregular outer boundaries if they are bounded
by ionization fronts. By means of models with clumped density
distributions, we have investigated the effects of the complex shapes
of the outer boundary on the analysis of ionic and elemental
abundances. We feel that our models contain the elements of a basic
physical effect that pertains to real H\,II regions in all of their
geometrical complexities: their boundaries are also complex in shape,
more so than our models.

We considered \citet{k94} stellar temperatures in the range 37.5 kK --
45 kK and ionizing photon luminosities of 10$^{48}$ -- 10$^{51}$
s$^{-1}$, all for an averaged density of 100 cm$^{-3}$. Of course,
other models of stellar atmospheres have been proposed. The most
important parameter of any of them is $\gamma$, the ratio of
He-ionizing photons to H-ionizing. We considered $\gamma$ in the range
0.57 -- 2.4. Our results apply rather well for densities $n<10^{4} $
cm$^{-3}$ to objects with the same $L_{49}/n$ as our models. With our
limited spatial resolution we found no regions truly shadowed from the
central star. Clumping makes significant differences on the true
abundances of ions of N, O, Ne, and S, as well as those derived from
emission lines plus measured nebular temperatures,
\toiii{} and \tnii. We considered two extreme limits of spatial
resolution: either none, so that only the fluxes of the emission lines
were used, or averaging the abundances as determined pixel-by-pixel
over the face of the nebula. There were no major differences
introduced by the two methods of averaging.

If \tnii{} is available for analyzing the low stages of ionization,
the overall abundances of O, Ne, and S are reasonably accurate because
of partially compensating errors.  For cool stars (for which most Ne is
Ne$^+$) in clumped models, (Ne/H) is badly estimated
from [Ne\,III] optical lines. All of these elements are well estimated
for uniform models. However, using \toiii{} for both low and high
stages of ionization leads to overestimates of (O/H) of $\sim$40 --
60\% for both clumped and uniform models, for cool stars.

The underlying physical cause of the differences between clumped and
uniform models is the flow of He recombination radiation from zones of
high ionization of H that can be in physical proximity to regions near
the outer edge of the nebula, with relatively large amounts of H$^0$.
We discuss this effect in \S3.1 and for each ion.

JSM acknowledges support from a PPARC Short Term Visiting Fellowship
to St Andrews, and KW from a PPARC Advanced Fellowship. Many of the
models were produced with computing facilities granted to Prof. Ellen
Zweibel by the Graduate School of the University of Wisconsin-Madison.
The comments of an anonymous referee improved the paper.

\label{lastpage} 

\begin{thebibliography}{}

\bibitem[\protect\citeauthoryear{Ackerman et al.}{2004}]{aetal04}
Ackerman, C. J., Carigi, L., Nissen, P. E., Pettini, M., \& Asplund, M.,
2004, A\&A, 414, 931

\bibitem[\protect\citeauthoryear{Ali et al.}{1991}]{a91}Ali, B., Blum,
R. D., Bumgardner, T. E., Cranmer, S. R., Ferland, G. J., Haefner, R.
I.,\& Tiede, G. P., 1991, PASP, 103, 1182

\bibitem[\protect\citeauthoryear{Asplund et al.}{Asplund et al.}
{2005}]{a04}Asplund, M., Grevesse, N., Sauval, A. J., Allende Prieto,
C., \& Blomme, R., 2005, A\&A, 431, 693

\bibitem[\protect\citeauthoryear{Draine}{Draine}{2004}]{d04}
Draine, B. T., 2004, web page at www.astro.princeton.edu/~draine

\bibitem[\protect\citeauthoryear{Elmegreen}
{Elmegreen}{1997}]{elm97} Elmegreen, B., 1997, ApJ, 477, 196

\bibitem[\protect\citeauthoryear{Esteban et al.}
{Esteban et al.}{1999a}]{em8} Esteban, C., Peimbert, M.,
Torres-Peimbert, S., Garc\'ia-Rojas, J., Ruiz, M. T.,
\& Rodr\'iguez, M., 1999, ApJS, 120, 113

\bibitem[\protect\citeauthoryear{Esteban et al.}
{Esteban et al.}{1999b}]{em17} Esteban, C., Peimbert, M.,
Torres-Peimbert, S., \& Garc\'ia-Rojas, J., 1999, Rev. Mex. Astron.
Astrof., 35, 65

\bibitem[\protect\citeauthoryear{Esteban et al.}
{Esteban et al.}{2004a}]{eori04} Esteban, C., Peimbert, M.,
Garc\'ia-Rojas, J., Ruiz, M. T., Peimbert, A., \& Rodr\'iguez,
M., 2004, MNRAS. 355, 229

\bibitem[\protect\citeauthoryear{Esteban et al.}
{Esteban et al.}{2005}]{ehii04} Esteban, C., Garc\'ia-Rojas, J.,
Peimbert, M., Peimbert, A., Ruiz, M. T., Rodr\'iguez, M., \&Carigi,
L., 2005, A\&A, 431, 693

\bibitem[\protect\citeauthoryear{Ferland et al.}
{Ferland et al.}{1995}]{fer95} Ferland, G. J., et al., 1995, in
Williams, R., Livio, M., eds., STScI Symp. 8, ``The Analysis of
Emission Lines'', Cambridge Univ. Press, Cambridge, 83

\bibitem[\protect\citeauthoryear{Garc\'a-Rojas et al.}
{Garc\'ia-Rojas et al.}{2004}]{gr04} Garc\'ia-Rojas, J., Esteban, C.,
Peimbert, M, Rodr\'iguez, M., Ruiz, M. T., \& Peimbert, A., 2004,
ApJS, 153, 501

\bibitem[\protect\citeauthoryear{Grevesse \& Sauval}{Grevesse \& 
Sauval}{1998}]{gs98} Grevesse, N., \& Sauval, A. J., 1998,
Sp. Sci. Rev., 85, 161

\bibitem[\protect\citeauthoryear{Holweger}{Holweger}{2001}]{hol}
Holweger, H., 2001, in Joint SOHO/ACE workshop "Solar and Galactic
Composition". Edited by Robert F. Wimmer-Schweingruber. Publisher:
American Institute of Physics Conference proceedings vol. 598

\bibitem[\protect\citeauthoryear{Howk \& Savage}{Howk \& Savage}
{1999}]{hs99}Howk, J. C., \& Savage, B. D., 1999, ApJ, 517, 746

\bibitem[\protect\citeauthoryear{Johnson \& Kobulnicky}{Johnson
\& Kobulnicky}{2003}]{jk03} Johnson, K. E., \& Kobulnicki, H. A.,
2003, ApJ, 597, 923

\bibitem[\protect\citeauthoryear{Kurucz}
{Kurucz}{1994}]{k94} Kurucz, R. L., 1994, CD-Rom 19, Solar Abundance
Model Atmospheres

\bibitem[\protect\citeauthoryear{Kwitter \& Henry}
{Kwitter \& Henry}{2002}]{kw01} Kwitter, K. B., \& Henry,
R. B. C., 2001, ApJ, 562, 804

\bibitem[\protect\citeauthoryear{Lagache et al.}{Lagache et al.}
{2000}]{lag00} Lagache, g., Haffner, L. M., Reynolds, R. J., \& Tufte,
S. L., 2000, A\&A, 354,247

\bibitem[\protect\citeauthoryear{MacLow}
{MacLow}{2004}]{ml04} Mac Low, M.-M., 2004, ApSS. 289, 323

\bibitem[\protect\citeauthoryear{MacLow \& Klessen}
{MacLow \& Klessen}{2004}]{mlk04} MacLow, M.-M., \&
Klessen, R. S., 2004, Rev. Modern Physics, 76, 125

\bibitem[\protect\citeauthoryear{Mathis, Whitney, \& Wood}
{Mathis, Whitney, \& Wood}{2002}]{mww02} Mathis, J. S., Whitney,
B. A., \& Wood, K., 2002, ApJ, 574, 812

\bibitem[\protect\citeauthoryear{O'Dell, Peimbert, \& Peimbert}
{O'Dell et al.}{2003}]{opp03}O'Dell, C. R., Peimbert, M., \&
Peimbert, A., 2003, AJ, 125, 2590

\bibitem[\protect\citeauthoryear{Olive \& Skillman}{Olive \& Skillman}
{2004}]{os04}Olive, K. A., \& Skillman, E. D.,  2004, ApJ, 617, 29O

\bibitem[\protect\citeauthoryear{Osterbrock}
{Osterbrock}{1989}]{deo89} Osterbrock, D. E., 1989, Astrophysics of
Gaseous Nebulae and active Galactic Nuclei, University Science Books,
Mill Valley, CA

\bibitem[\protect\citeauthoryear{Pagel \& Tautvai\v{s}ien\.{e}}
{Pagel \& Tautvai\v{s}ien\.{e}}{1995}]{pt95}Pagel, B. E. J., \&
Tautvai\v{s}ien\.{e}, G., 1995, MNRAS, 276, 505

\bibitem[\protect\citeauthoryear{Peimbert et al.}
{Peimbert et al.}{2000}]{ppm00} Peimbert, M., Peimbert, A.,
\& Ruiz, M.-T., 2000, ApJ, 541, 688

\bibitem[\protect\citeauthoryear{P\'equignot, D., et al.}
{P\'equignot, D., et al.}{2001}]{peq01} P\'equignot, D., et al., 2001,
in ``Spectroscopic Challenges of Photoionized Plasmas'', G. Ferland \&
D. W. Savin, eds., ASP Conference Series 247. San Francisco: ASP, 533

\bibitem[\protect\citeauthoryear{Savage \& Sembach}
{Savage \& Sembach} {1996}]{ss96} Savage, B. D., \& Sembach,
K. R., 1996, ARAA, 34, 279

\bibitem[\protect\citeauthoryear{Sellgren et al.}{Sellgren et al.}
{1990}]{s90}Sellgren, K., Tokunage, A./ T., \& Nakada, Y., 2004, ApJ, 
349, 120

\bibitem[\protect\citeauthoryear{Sofia et al.}
{Sofia et al.}{2004}]{s04} Sofia, U. J., Laroesch, J. T., Meyer, D.,
\& Cartledge, S. I. B., 2004, ApJ, 605, 272

\bibitem[\protect\citeauthoryear{Tsamis et al.}
{Tsamis et al.}{2003}]{tea03}Tsamis, Y. G., Barlow, M. J., Liu, X.-W.,
Danziger, I. J., \& Storey, P. J., 2003, MNRAS, 338,687

\bibitem[\protect\citeauthoryear{Weingartner \& Draine}
{Weingartner \& Draine}{2001}]{wd01}Weingartner, J. C., \& Draine,
B. T., 2001, ApJ, 548, 296

\bibitem[\protect\citeauthoryear{Wood, Mathis, \& Ercolano}
{Wood et al.}{2004}]{wme04} Wood, K., Mathis, J. S., \& Ercolano,
B., 2004, MNRAS, 348, 1337

\bibitem[\protect\citeauthoryear{Zubko, Dwek, \& Arendt}{Zubko et al.}
{2004}]{zda04} Zubko, V. Dwek, E., \& Arendt, R. G., 2004, ApJS, 152, 211

\end{thebibliography}
\end{document}